# The many faces of the Bohr atom

*Helge Kragh*[*]

The atomic model that Bohr proposed in 1913 constituted a break with all earlier conceptions of the atom. Keeping to the theory's basic postulates – the stationary states and the frequency condition – he conceived the model as preliminary and immediately began developing and modifying it. Strictly speaking there was no single Bohr atom but rather a series of different models sharing some common features. In this paper I start with calling attention to some less well known aspects of Bohr's early model of one-electron atoms the significance of which only became recognized after his death in 1962. I then briefly sketch how he abandoned the ring model for many-electron atoms about 1920 and subsequently went on developing the ambitious orbital model that he thought would unravel the secrets of the periodic system. Bohr's model of 1921-1922 marked the culmination of the orbital atom within the old quantum theory, but it would soon be replaced by a symbolic and non-visualizable view of atomic structure leading to the atom of quantum mechanics.

## 1. Rydberg atoms and isotope effect

Among the unusual features of the atom that Bohr presented in the first part of his trilogy of 1913 was that the hydrogen atom, and other atoms as well, did not have a fixed size. For the radius of a one-electron atom with nuclear charge $Ze$ he derived the expression

$$a_\tau = \tau^2 a_1, \quad a_1 = \frac{h^2}{4\pi^2 Zme^2},$$

where $\tau = 1, 2, 3,\ldots$ and the other symbols have their usual meanings. For a hydrogen atom ($Z = 1$) in its ground state $\tau = 1$ he found the value $a_1 = 0.55 \times 10^{-8}$ cm as compared to the modern value $0.53 \times 10^{-8}$ cm. Known as the Bohr radius, the quantity is usually designated the symbol $a_0$ rather than $a_1$. As Bohr pointed out later in his paper, because the radius of the atom varies as $\tau^2$, an atom in a highly excited state can be remarkably large: "For $\tau = 12$ the diameter is equal to $1.6 \times 10^{-6}$ cm, or equal to the mean distance between the molecules in a gas at a pressure of about 7 mm mercury; for $\tau = 33$ the diameter is equal to $1.2 \times 10^{-5}$ cm, corresponding to the mean distance between the molecules at a pressure of about 0.02 mm mercury."[1] Because the area of the electron's orbit varies as $\tau^4$, for $\tau$ = 33 it is about a million times the area in an ordinary atom.

---

[*] Centre for Science Studies, Department of Physics and Astronomy, Aarhus University. E-mail: helge.kragh@ivs.au.dk.

Bohr used his observation to come up with a clever explanation of why only 12 of the Balmer lines had been found in experiments with vacuum tubes, while astronomers had observed as many as 33 lines. He argued that spectral lines arising from hydrogen atoms in high quantum states required a very low density, although "for simultaneously to obtain an intensity sufficient for observation the space filled with the gas must be very great." These conditions he thought might exist only in the rarefied atmosphere of the stars. "We may therefore never expect to be able in experiments with vacuum tubes to observe the lines corresponding to high numbers of the Balmer series of the emission spectrum of hydrogen."[2] Bohr returned to the question in his section on the absorption of radiation, where he discussed experiments made by the American physicist Robert Wood on absorption of light by sodium vapour. Wood had observed about 50 absorption lines and that although the pressure was not very low. This Bohr explained as "an absorption of radiation which is not accompanied by a complete transition between two stationary states … [but] is followed by an emission of energy during which the systems pass back to the original stationary state."[3]

The kind of monster-atoms introduced by Bohr was later called "Rydberg atoms" because the frequencies are included in the spectral formula that Janne Rydberg proposed in 1890. Since Rydberg's work was purely empirical, "Rydberg-Bohr atoms" might seem to be a more appropriate name.[4] In 1913 Bohr derived the quantized energy levels of hydrogen by considering quantum jumps between adjacent states at very high values of $\tau$, where the separation of adjacent energy levels varies as $1/\tau^3$. For the ratio of the mechanical frequencies of the states $\tau = N$ and $\tau = N - 1$ he found

$$\frac{\omega_N}{\omega_{N-1}} = \frac{(N-1)^3}{N^3},$$

---

[1] Bohr 1913a, p. 9.
[2] Ibid., p. 10.
[3] Ibid., p. 18.
[4] The term "Rydberg atom" only came into wide use in the late 1970s. According to the *Web of Science*, it was first used in the title of a scientific paper in 1971. Kleppner, Littman and Zimmerman 1981 emphasize Bohr's role as the founding father of the physics of Rydberg atoms.

which tends toward unity for $N \gg 1$. Bohr used the result to argue that for highly excited states the radiation frequency due to quantum jumps would be almost the same as the mechanical frequency of revolution. Thus, in this first discussion of the correspondence principle for frequencies he was in effect using Rydberg states as an illustration.

Bohr's expectation with regard to highly excited atoms turned out to be basically correct. Isolated Rydberg atoms were first observed deep in interstellar space.[5] In 1965 scientists from the National Radio Astronomy Observatory in the USA detected microwave radiation from hydrogen atoms corresponding to transitions between energy levels near $\tau = 100$, and later radio astronomers have detected states as large as $\tau = 350$ in outer space. Because of the exceedingly low density in interstellar gas clouds, Rydberg atoms can exist for long periods of time without being ionized. Whereas the life-time of an ordinary excited atom is of the order $10^{-8}$ second, Rydberg atoms may live as long as a second. Astronomers have for long been familiar with a radiation from the heavens at a frequency of 2.4 GHz that is due to a transition in hydrogen from $\tau = 109$ to $\tau = 108$. It was only with the arrival of tunable dye lasers in the 1970s that it became possible to study Rydberg atoms in the laboratory, after which the subject became increasingly popular. Today it has grown into a minor industry.[6]

After Bohr had presented his atomic theory in the July 1913 issue of *Philosophical Magazine*, the British spectroscopist Alfred Fowler objected that Bohr's theoretical wavelengths for hydrogen and the helium ion He⁺ did not agree precisely with those found experimentally. According to Bohr's theory,

$$\frac{1}{\lambda} = Z^2 R_\text{H} \left( \frac{1}{\tau_2^2} - \frac{1}{\tau_1^2} \right), \quad \text{with} \ \ R_\text{H} = \frac{2\pi^2 m e^4}{h^3 c}$$

As well known, Bohr responded to Fowler's challenge by taking into account the finite mass of the nucleus, namely, by replacing the electron mass *m* by the reduced mass given by

$$\mu = \frac{mM}{m + M} = \frac{m}{1 + m/M},$$

---

[5] Dalgarno 1983.
[6] See Gallagher 1994 for a comprehensive but largely non-historical review of Rydberg atoms. There is no scholarly study of the history of the subject.

where *M* denotes the mass of the nucleus. In this way the Rydberg constant would depend slightly on *M*, causing the discrepancies mentioned by Fowler to disappear.[7] For an infinitely heavy nucleus ($m/M = 0$), Bohr calculated

$$R_\infty = R_H \left(1 + \frac{m}{M_H}\right) = 109\,735 \text{ cm}^{-1}$$

where $M_H$ is the mass of a hydrogen nucleus (Figure 1). In his Bakerian Lecture of 1914, Fowler used Bohr's expression to derive a mass ratio of the hydrogen nucleus (proton) and the electron of $M_H/m = 1836 \pm 12$.[8]

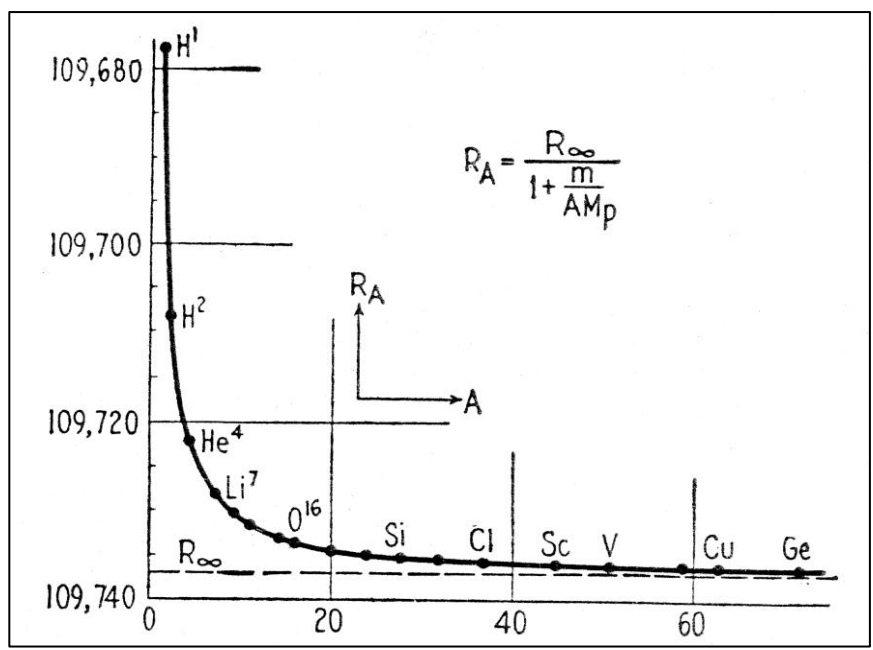

Figure 1. Variation of Rydberg's constant with the atomic weight *A*. Source: Harvey E. White, *Introduction to Atomic Spectra* (New York: McGraw-Hill, 1934), p. 37.

In late 1913 Bohr realized that the introduction of the reduced mass would result in a small shift between the lines of two isotopic atoms. As early as September 1913, during the Birmingham meeting of the British Association for the Advancement of Science, he had suggested that the positive rays that J. J. Thomson ascribed to the $H_3$ molecule might be due to a hydrogen isotope of mass 3 (that is, $^3H^+$ instead of Thomson's $H_3^+$).[9] While he did not think of a spectral shift as evidence for the hypothesis in Birmingham, this is what he did a few months later. Together with his colleague in Copenhagen, the spectroscopist Hans Marius Hansen, he even conducted experiments to detect the H-3 isotope later known as tritium.[10] It follows from Bohr's theory that the isotope shift would be

$$\lambda_T - \lambda_H = \left(1 - \frac{R_H}{R_T}\right)\lambda_H = \frac{2m}{3m + 3M_H}\lambda_H = 3.6 \times 10^{-4}\lambda_H$$

---

[7] Bohr 1913b.
[8] Fowler 1914. The presently known proton-to-electron mass ratio is 1836.3.
[9] For the story of triatomic hydrogen and references to the literature, see Kragh 2012a.

Bohr continued for some time to think of the isotope shift. Although he did not refer to it in his publications, he mentioned it at the September 1915 meeting of the British Association in Manchester, from where it found its way into *Nature*.[11] Apparently he did not consider it very important.

The spectroscopic isotope effect was discovered in molecules in 1920, independently by Francis Loomis in the United States and Adolf Kratzer in Germany who were both able to separate the vibrational frequencies in HCl due to the isotopes Cl-35 and Cl-37. It took another twelve years until the corresponding atomic effect was confirmed. In 1913 Bohr had contemplated the existence of H-3 but not the isotope of mass 2 that Harold Urey, George Murphy and Ferdinand Brickwedde detected spectroscopically in 1932. The Nobel Prize-rewarded discovery of deuterium was directly guided by Bohr's old theory of the isotope effect. This effect later became very important as a method applicable to a variety of sciences ranging from physics and chemistry over astronomy to geology and biology.[12]

## 2. Many-electron atoms

Bohr's theory of 1913 was much more than just a theory of the hydrogen atom. In the second part of the trilogy he ambitiously proposed models also of the heavier atoms, picturing them as planar systems of electrons revolving around the nucleus. The lithium atom, for example, would consist of two concentric rings, an inner one with two oppositely located electrons and an outer one with a single electron. Ring structures of this kind had already been proposed by Thomson in his older atomic theory, and Bohr relied to some extent on Thomson's method with regard to calculations of mechanical stability. Bohr's ring-atoms were soon developed by Walther Kossel, Arthur Compton, Peter Debye, Lars Vegard and other physicists who for a while thought that the model was supported by X-ray spectroscopic data. However, latest by 1920 it was realized that the planar ring atom was inadequate and had to be replaced by a more complex model that made both chemical and physical sense.[13]

---

[10] See Kragh 2012b, pp. 97-98 and Kragh 2012c. Tritium does not exist naturally. It was first produced in nuclear reactions in 1939.
[11] *Nature* 96 (1915), 240.
[12] Wolfsberg, Van Hook and Paneth 2010. For the discovery of deuterium, see Brickwedde 1982.

In a lecture of 1921 the American physical chemist Richard Tolman criticized the physicists' "absurd atom, like a pan-cake of rotating electrons" and their naïve picture of the carbon atom "as a positive nucleus with rings of electrons rotating around it in a single plane."[14] However, this was no longer Bohr's view. Neither was it a view shared by the majority of quantum physicists, who by then had arrived at the conclusion that the atom must have a spatial architecture. As an extension of Arnold Sommerfeld's *Ellipsenverein*-model of 1918, Alfred Landé developed a class of models governed by cubical and other polyhedral symmetries, what he called *Würfelatome*. Landé's cubical atoms, with electrons moving in small orbits at the corners of concentric cubes, were seen as a welcome break with the planar atom and for this reason received positively in the physics community. Bohr found Landé's ideas to be of such interest that he invited him to Copenhagen. However, when Landé gave his lecture in Copenhagen in October 1920, Bohr had reached the conclusion that the cubical atom was not the answer to the puzzle of the complex atoms.

Bohr agreed that the simple ring atom had to be abandoned, such as he wrote to Owen Richardson on Christmas day 1919: "I am quite prepared, or rather more than prepared, to give up all ideas of electronic arrangements in 'rings'."[15] Half a year later he was working on a new picture of the atom as consisting of spatially structured elliptical orbits. There is little doubt that the new picture was in part motivated by the unsatisfactory calculations that he, together with his assistant Hendrik Kramers, had performed in order to understand the helium atom. Bohr and Kramers came to the conclusion that the ground state of helium could not be represented by a planar structure but more likely be pictured as two intersecting circular orbits (Figure 2).

In a letter to Rudolf Ladenburg of 16 July 1920 Bohr wrote, "it also seems that an assumption of rings already has to be given up because of insufficient stability and that we are forced to expect much more complicated motions of the electrons in the atoms."[16] What these more complicated motions were he first revealed in a lecture to the Physical Society in Copenhagen on 15 December 1920. A published version of his new ideas only appeared in March 1921, in an unusually long communication to the letters section of *Nature*.

---

[13] See Heilbron 1967 and Kragh 2012b, which include references to the literature.
[14] Tolman 1922, p. 222 and p. 226. The pancake metaphor was also used by Sommerfeld, see his letters to Landé as quoted in Heilbron 1967, p. 479.
[15] Bohr to Richardson, 25 December 1919, as quoted in Heilbron 1967, p. 478.

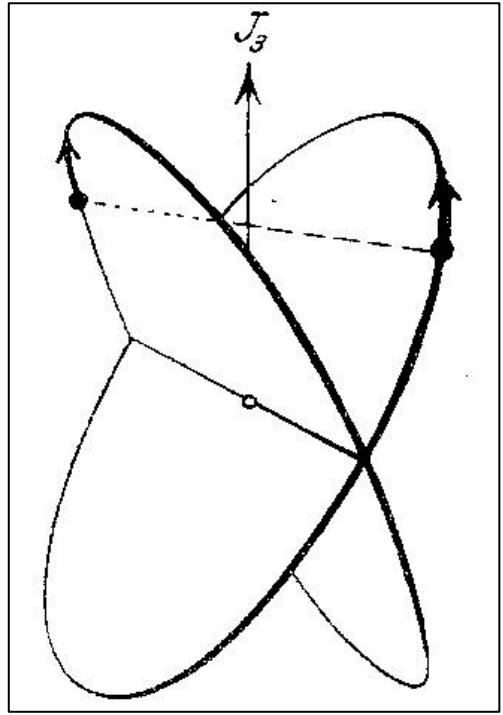

Figure 2. The helium atom with two crossed orbits according to Bohr and Kramers. Source: Max Born, *Vorlesungen über Atommechanik* (Berlin: Julius Springer, 1925), p. 331.

## 3. The final orbital atom

In the period from about 1921 to 1924, Bohr's new model of complex atoms, or of the periodic system of the elements, was widely discussed and acclaimed. Sommerfeld's response to Bohr's first announcement of his theory – that "it evidently represents the greatest advance in atomic structure since 1913" – was echoed by many of his colleagues in atomic and quantum physics.[17] Bohr himself held great hope in his theory which he developed in several papers and discussed in high-profile lectures such as the Wolfkehl lectures in Göttingen in June 1922 and the Nobel lecture in Stockholm six months later. He elaborated his original version in another letter to

---

[16] Quoted in Kragh 2012b, p. 272.
[17] Sommerfeld to Bohr, 7 March 1921, quoted in Kragh 2012b, p. 298.

*Nature* of September 1921 and, in great detail, in an extended published version of a lecture he gave in Copenhagen the following month. "The Structure of the Atom and the Physical and Chemical Properties of the Elements" gained a wide readership and convinced many physicists that Bohr's approach was the key to unlock the secrets of the atom.[18] Apart from Bohr's own writings and lectures, the theory appeared prominently in scientific as well as popular books. Sommerfeld dealt with it in his *Atombau und Spektrallinien* (1922, 1924), Max Born in his *Vorlesungen über Atommechanik* (1925), and Kramers and Helge Holst in their popular book *The Atom and the Bohr Theory of its Structure* (1923).

Rather than describing the historical development and reception of the theory, I shall summarize its basic features and methodological foundation in the form the theory was known in early 1923.[19] Bohr adopted the extension of his original theory that Sommerfeld had proposed in 1915 by replacing the circular electron orbits with elliptical orbits. Following Sommerfeld, he specified the orbit of an electron by its principal quantum number $n$ and its azimuthal quantum number $k$, the two numbers attaining values given by $n = 1, 2, 3, …$ and $k = 1, 2, … n$. In the case of $k = n$ the orbit is circular, whereas $k < n$ indicates a Kepler ellipse whose eccentricity increases with $n – k$. To build up a theory on this basis, Bohr relied on two hypotheses, the *Aufbau* principle and the penetration effect. By using these and other hypotheses he hoped to explain the so-called Rydberg rule, according to which the number $N$ of elements in the various periods can be written as

$$N = 2n^2, \qquad n = 1, 2, 3, …$$

Bohr considered the structure of a neutral atom to be the result of how it was formed by the successive addition of $Z$ electrons to a bare nucleus. According to the so-called *Aufbauprinzip* (construction or building-up principle), the addition of electron number $p$ to a partially completed atom with $p – 1$ bound electrons would leave the quantum numbers of the $p – 1$ electrons unchanged. When, in this building-up process, a new atom is formed, the principal quantum number of the last captured electron will differ from that of the already bound electrons in the outer shell only if the atom being formed belongs to a new period of the periodic system. Thus, in each new period $n$ increases by one unit.

---

[18] Bohr 1922, with translations into English, French and Russian.
[19] For details, see Kragh 1979 and Kragh 2012b, pp. 271-302.

In order to explain the finer details of the periodic system, and especially the transition groups and the rare earths, Bohr made use of the hypothesis of penetrating orbits, which was essential to his entire line of argument. According to this hypothesis, the valence or optical electrons moving in eccentric elliptical orbits would penetrate the inner shell of eight electrons that characterize the noble gases. The idea of penetrating orbits was independently suggested by Erwin Schrödinger in 1921, but only Bohr applied it systematically to the periodic system. In Bohr's theory, the penetrating orbits not only accounted for the spectra of the alkali metals, but above all they played the role of a coupling effect. He pictured the penetrating orbits as divided in two parts: outside the core of the atom the optical electron moves in an approximate Keplerian ellipse exhibiting a perihelion precession; when it penetrates the region of the core, the electric field is changed and the internal orbit is no longer a simple continuation of the outer elliptical orbit. Instead, it performs an orbit much closer to the nucleus and therefore is more effectively bound.

From a methodological point of view, Bohr's theory was markedly eclectic, relying on a peculiar mixture of empirical evidence and theoretical reasoning (Figure 3). Among the empirical evidence were data from X-ray spectroscopy, which he investigated together with the Dutch physicist Dirk Coster, but these played no role in his original formulation of the theory.

The two components were tied together by a more intuitive understanding of the mechanism in the building up of atoms. Although the new theory was no less dependent on empirical knowledge of the chemical elements than the 1913 theory, Bohr stressed that it was not derived inductively from such knowledge. It was the use of general principles that distinguished the new theory from earlier ideas of atomic structure and supplied it with Bohr's personal imprint. These general principles he used in a philosophical rather than physical or mathematical way, in the sense that they were not stated quantitatively but were qualitative considerations of an intuitive kind. Foremost among them was the versatile correspondence principle which permeated the entire theory in a characteristic but also opaque way. For example, he claimed that the quantum state of an atom could be inferred from a "closer investigation" based on the correspondence principle. Bohr's "closer investigation" – a favourite phrase in his idiosyncratic terminology – remained unclear, except that it did not imply a mathematical deduction from quantum theory or, for that matter, calculations at all.

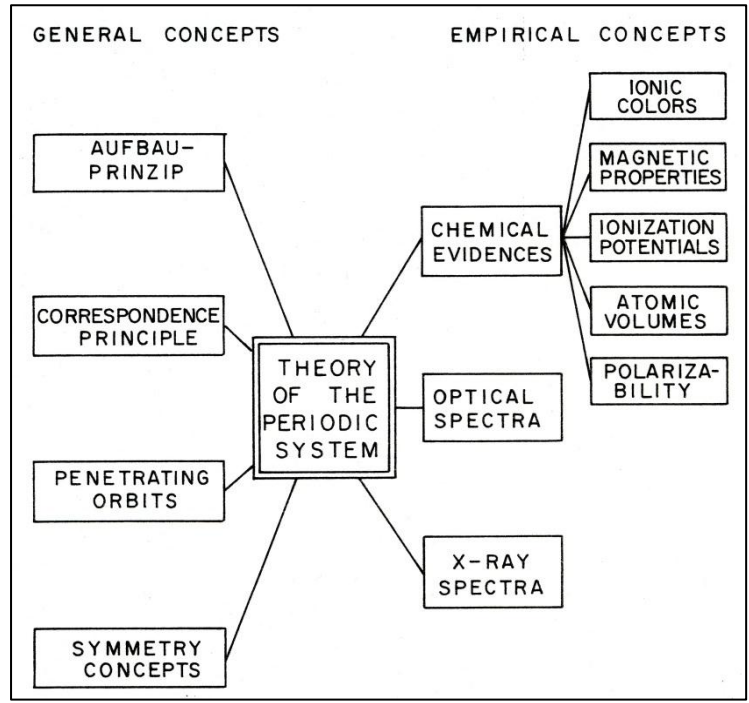

Figure 3. The conceptual structure of Bohr's 1922 theory of atomic structure.
Source: Kragh 1979, p. 145.

More often than not, the correspondence principle acted as a *deus ex machina*, or so it seemed to many contemporary physicists outside the Copenhagen group. Kramers, who knew Bohr's style of physics intimately, recalled how he had arrived at his theory: "It is interesting to recollect how many physicists abroad thought, at the time of the appearance of Bohr's theory of the periodic system, that it was extensively supported by unpublished calculations which dealt in detail with the structure of the individual atoms, whereas the truth was, in fact, that Bohr had created and elaborated with a divine glance a synthesis between results of a spectroscopical nature and of a chemical nature."[20]

## 4. The orbital atom dismounted

The result of Bohr's elaborate considerations was a picture of the atom as consisting of electrons moving in an interlocked, harmonious system of elliptical orbits with different eccentricities. Because of the slow precession of the ellipses, the orbits would not be closed but slightly open. Moreover, some of the orbits would penetrate into the inner electronic system and therefore change their form. In the plates that Bohr used for his lectures and which were reproduced in Kramers and Holst's book on his atomic theory, the atoms were shown in two dimensions and with all the orbits roughly drawn to scale (Figure 4). In reality the electron orbits made up a three-dimensional structure. Is this what Bohr thought an atom looked like? Did he consider the picture as a realistic or merely a symbolic representation of the atom?

---

[20] Kramers 1935, p. 90.

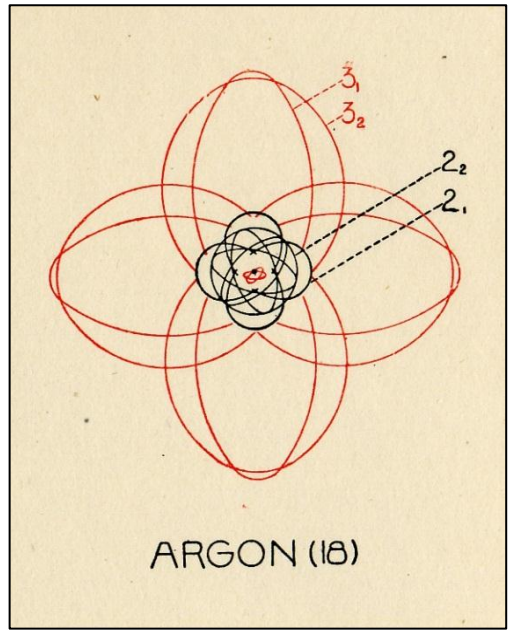

Figure 4. Bohr's symmetric structure of the argon atom, showing its orbitals denoted as $n_k$, where $n$ is the principal and $k$ the azimuthal quantum number. Source: Kramers and Holst 1923, plate II.

      Commenting on their diagrams, Kramers and Holst warned that, "Although the attempt has been made to give a true picture of these orbits as regards their dimensions, the drawings must still be considered as largely symbolic."[21] Also Bohr seems to have believed that pictures of the atom should be understood as analogies or symbols. In a letter of 22 September 1922 to the Danish philosopher Harald Høffding, Bohr expressed his doubts "that we shall be able, in the world of the atom, to carry through a description in space and time of a kind which corresponds to our ordinary sensory image." He stressed that "one is operating with analogies."[22] On the other hand, at the time neither Bohr nor Kramers had apparently any doubt about

---

[21] Kramers and Holst 1923, p. 192. On this book, see Kragh and Nielsen 2012.
[22] Quoted in Kragh 2012b, pp. 352-353.

the reality of electron orbits or the fertility of the semimechanical model concept on which Bohr's new theory rested. Although the atom did not quite look like the picture, it might still be something like it. Bohr may not have thought of his atom as something corresponding to "our ordinary sensory image," and yet this was the impression his lectures and articles conveyed to most of his colleagues in physics.

Bohr's atomic theory of the periodic system was short-lived. It was soon replaced by Pauli's theory based on the exclusion principle, which stood in stark contrast to Bohr's. Pauli not only denied the validity of the correspondence principle in building up atomic structures, he also rejected the concept of electron orbits. To Bohr he wrote: "I have avoided the term 'orbit' altogether in my paper … I think the energy and [angular] momentum values of the stationary states are something much more real than the 'orbits'."[23]

By the summer of 1924 the kind of visualizable model that characterized Bohr's theory was no longer considered a candidate for the real structure of atoms. Objections to the orbital model had been around for some time, raised in particular by the youngsters Pauli and Heisenberg, whereas it took more time for Bohr to abandon the orbits.[24] Still in the autumn of 1923 he maintained orbits in the stationary states, although no longer governed by the rules of classical mechanics. In the Bohr-Kramers-Slater (BKS) theory from 1924, describing the atom as an orchestra of virtual oscillators, the electrons orbiting in stationary states had finally disappeared. To the extent that the atom of the BKS theory can be called a model at all, it was entirely different from the pictorial model that Bohr had introduced with such confidence just three years earlier. This was even more the case with the symbolic model of the atom that Heisenberg proposed in the summer of 1925 and which marked the beginning of quantum mechanics.

---

[23] Pauli to Bohr, 12 December 1924, quoted in Kragh 2012b, p. 307. See also Heilbron 1983.
[24] Darrigol 1992, p. 167 and p. 199, argues that Bohr, as early as March 1922, considered electron orbits to be of a formal nature only.